\documentclass{ws-procs9x6}

\begin{document}

\title{Deformations and quasiparticle spectra of nuclei in the nobelium
region}

\author{Yue Shi,$^{1,2,3}$ J. Dobaczewski,$^{3,4}$
P.T. Greenlees,$^{3}$ J. Toivanen,$^{3}$
and P.~Toivanen,$^{3}$}

\address{
$^{1}$Department of Physics and Astronomy, University of
Tennessee, Knoxville, Tennessee 37996, USA\\
$^{2}$Physics Division, Oak Ridge National Laboratory, Post
Office Box 2008, Oak Ridge, Tennessee 37831, USA\\
$^{3}$Department of Physics, PO Box 35 (YFL), FI-40014
University of Jyv{\"a}skyl{\"a}, Finland\\
$^{4}$Institute of Theoretical Physics, Faculty of Physics,
University of Warsaw, ul. Ho{\.z}a 69, PL-00681 Warsaw, Poland
}

\begin{abstract}
We have performed self-consistent Skyrme Hartree-Fock-Bogolyubov
calculations for nuclei close to $^{254}$No. Self-consistent
deformations, including $\beta_{2,4,6,8}$ as functions of the
rotational frequency, were determined for even-even nuclei
$^{246,248,250}$Fm, $^{252,254}$No, and $^{256}$Rf. The quasiparticle
spectra for $N=151$ isotones and $Z=99$ isotopes were calculated and
compared with experimental data and the results of
Woods-Saxon calculations.
We found that our calculations give high-order deformations
similar to those
obtained for the Woods-Saxon potential, and that the experimental
quasiparticle energies are reasonably well reproduced.
\end{abstract}

\keywords{Self-consistent mean-field calculations;
self-consistent deformations; quasiparticle energies.}

\bodymatter

\section{Introduction}\label{intro}

Nuclei in the nobelium region represent the heaviest systems for which
detailed spectroscopic information on the structure is
available~\cite{herzberg08}. Hence, they provide an excellent testing
ground for various theoretical models that aim at reliable
predictions of properties of superheavy elements. Although
significant progress has been made both experimentally and
theoretically, there are still challenges for self-consistent
mean-field calculations (see Ref.~\refcite{bender12} for a recent
review). One of the challenges is the correct description of deformed shell gaps.
Macroscopic-microscopic calculations predict shell gaps at $N=152$
and $Z=100$, whereas typical self-consistent mean-field calculations give
shell gaps at $N=150$, and $Z=98$ and $Z=104$. The experimental findings
are consistent with the macroscopic-microscopic shell gaps.

Current advanced readjustements~\cite{kortelainen12} of coupling
constants of the Skyrme functional do not cure this deficiency,
although they point to a large underdetermination of the spin-orbit
properties. To study detailed spectroscopic properties of heavy and
superheavy nuclei, one thus may attempt a fine tuning of the
spin-orbit coupling constants, so as to bring the deformed shell gaps
in closer agreement with data. For this purpose, we recently
readjusted~\cite{shi13} the spin-orbit coupling constants,
$C_0^{\nabla J}$ and $C_1^{\nabla J}$, and pairing strengths,
$V_0^{\nu}$ and $V_0^{\pi}$, of the Skyrme UNEDF1
functional~\cite{kortelainen12} to match the experimental excitation
energies of the $11/2^-$ state in $^{251}$Cf and $7/2^+$ state in
$^{249}$Bk.
For the time-odd coupling constants of the functional, we adopted the
prescription based on the Landau parameters, as defined in Ref.~\refcite{[Ben02]}.
The remaining parameters of UNEDF1 were kept unchanged.
We call this new parameter set UNEDF1$^{\textrm{SO}}_L$.
Details of the adjustment procedures will be presented in
a forth-coming publication~\cite{shi13}. Here we only quote the obtained values,
which read,
\begin{eqnarray}
(C_0^{\nabla J}, C_1^{\nabla J}) &=& (-88.05040, 8.45838)\,\mbox{MeV\,fm$^5$},\\
(V_0^{\nu}, V_0^{\pi}) &=& (-205.05, -252.48)\,\mbox{MeV}.
\end{eqnarray}

As a result, in Ref.~\refcite{shi13} we found improvements of shell
gaps, quasiparticle energies, and moments of inertia (MoI) in nuclei
around $^{254}$No. Also a good agreement with experiment has been
obtained for the quasiparticle energies in $^{249}$Bk and $^{251}$Cf.
The single-particle spectra more closely resemble those obtained with
the WS potential, and shell gaps are opened at $Z=100$ and $N=152$ in $^{254}$No.
In the present work, we extended the calculations to other
nuclei in the nobelium region. We investigated the self-consistent
deformations as a function of frequency and we performed blocked
calculations for odd-$A$ $N=151$ isotones and Es ($Z=99$) isotopes.

\section{The model}\label{model}

In this work, all the calculations were performed by using the
symmetry-unrestricted solver HFODD~\cite{schunck12} (v2.52j).
For rotational calculations, the standard cranking term $-\omega J_y$ was added to the
Hamiltonian. The effect of an odd nucleon was treated by employing the blocking
approximation, that is, by using the density matrix and
pairing tensor in the form~\cite{ring80},
\begin{eqnarray}
\rho_{mn}^{\mu} &=& (V^*V^T)_{mn} + U_{m\mu}U^*_{n\mu} - V^*_{m\mu}V_{n\mu},\\
\kappa_{mn}^{\mu} &=& (V^*U^T)_{mn} + U_{m\mu}V^*_{n\mu} - V^*_{m\mu}U_{n\mu}.
\end{eqnarray}
We refer the reader to Ref.~\refcite{schunck10} for details.

In the cranking-plus-blocking calculations, the time-reversal symmetry is broken, so
the time-odd terms of the Skyrme density functional are present.
The importance of time-odd terms in describing collective rotational bands has
been emphasized in Ref.~\refcite{dobaczewski95}.
The advantage of using a symmetry-unrestricted solver is that the effects of non-zero
time-odd potentials can be taken into account in a strict manner.

We used 680 deformed harmonic-oscillator basis states, with the
basis-deformation parameters of $\hbar \omega_x$ = $\hbar \omega_y$ =
8.5049\,MeV and $\hbar \omega_z$ = 6.4823\,MeV. The cutoff energy in
the quasiparticle spectrum was $E_c=60$\,MeV. In the pairing channel,
to describe the approximate particle-number projection, we adopted
the Lipkin-Nogami formalism.

\section{Results}

In the nobelium region, high-order deformations of the WS
potential were found to play important roles in explaining many
high-spin phenomena. For example, the decrease of $E_x(2^+)$ in
$^{254}$No~\cite{sobiczewski01} and the delayed upbending of MoI in
$^{254}$No as compared to $^{252}$No~\cite{liu12} were attributed to
the effect of $\beta_6$ deformation on the relevant single-particle levels.

We performed cranking calculations for the 6 nuclei, in which the
ground-state rotational bands are observed, namely, for
$^{246,248,250}$Fm, $^{252,254}$No, and $^{256}$Rf. We determined the
MoIs, proton and neutron pairing gaps, and self-consistent
deformations as functions of the rotational frequency~\cite{shi13}.
The experimental kinematic and dynamic MoIs were well reproduced by
our calculations. Deformations obtained within our fully
self-consistent calculations provide a reasonable description of
polarization effects exerted by the occupations of individual
deformed orbitals. They were determined as the Bohr deformation
parameters $\beta_\lambda$ of sharp-edge mass distributions that have
multipole moments $Q_{\lambda}$, for $\lambda=2$--8, identical to
those calculated microscopically, see Ref.~\refcite{dobaczewski09}.
All nuclei considered here have axial ground-state shapes.

\begin{figure}
\begin{center}
\psfig{file=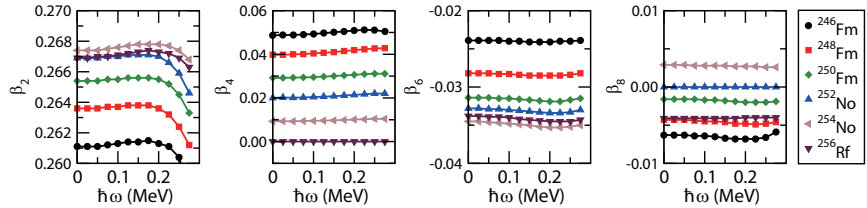,width=4.5in}
\end{center}
\caption{Deformations $\beta_{2,4,6,8}$ as functions of the
rotational frequency, calculated self-consistently for six nuclei,
$^{246,248,250}$Fm, $^{252,254}$No, and $^{256}$Rf.}
\label{fig1}
\end{figure}

Figure~\ref{fig1} shows deformations $\beta_{2,4,6,8}$ as a function
of rotational frequency $\omega$. At $\omega=0$,
deformations $\beta_2$ are systematically higher than those obtained by
macroscopic-microscopic WS calculations~\cite{sobiczewski01,liu12},
and, in general, lower than those obtained within other relativistic and
non-relativistic mean-field calculations~\cite{afanasjev03,bender03}. It
is interesting to note that the high-order deformations presented in
Figure~\ref{fig1} are very close to those of
Refs.~\refcite{sobiczewski01} and~\refcite{liu12}. With increasing rotational
frequency, the deformations stay almost constant. Only at $\omega \sim
0.3$~MeV, deformations $\beta_2$ begin to decrease. This behavior
is consistent with that of Ref.~\refcite{liu12}.

\begin{figure}
\begin{center}
\psfig{file=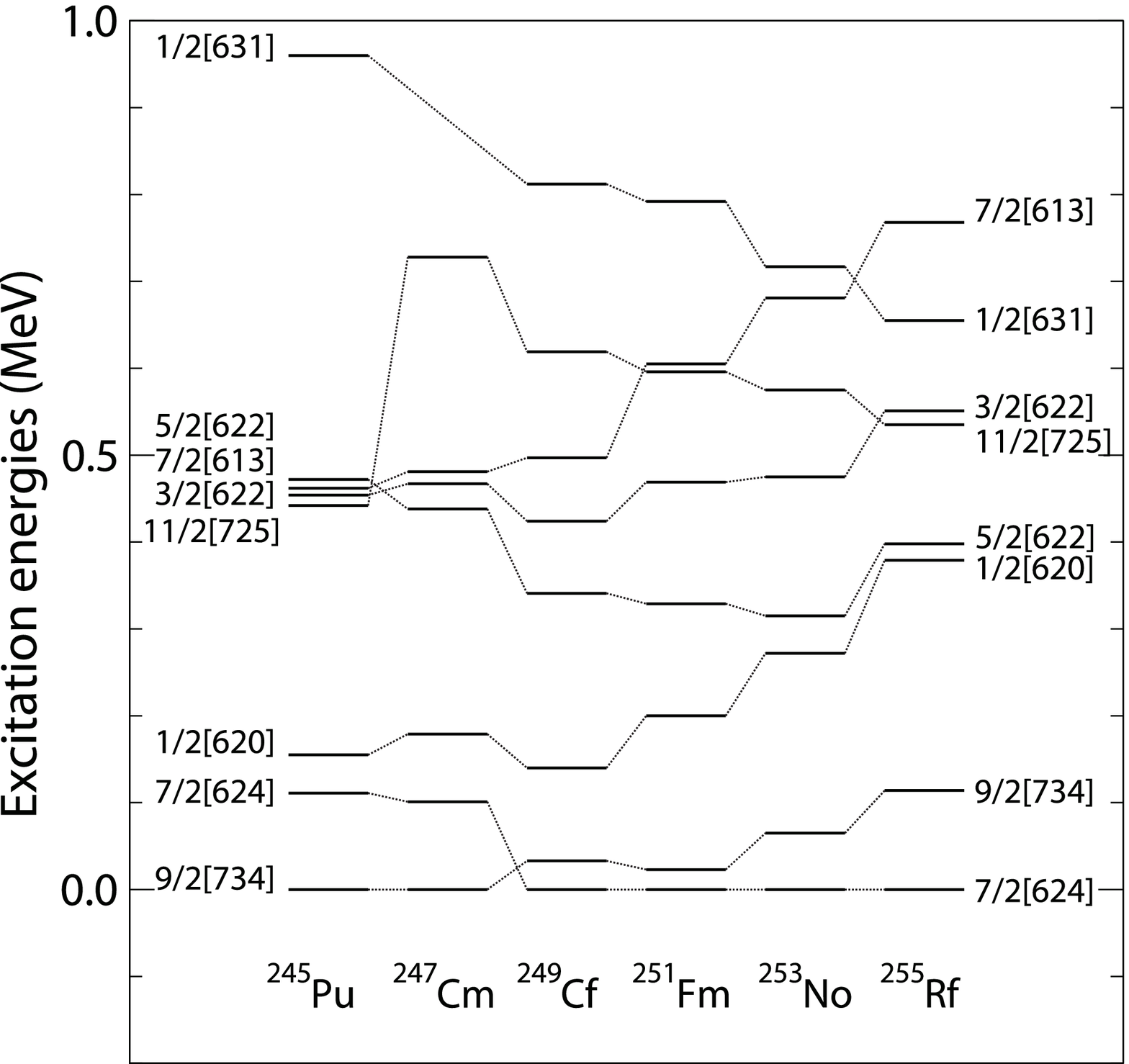,width=2.2in}~~%
\psfig{file=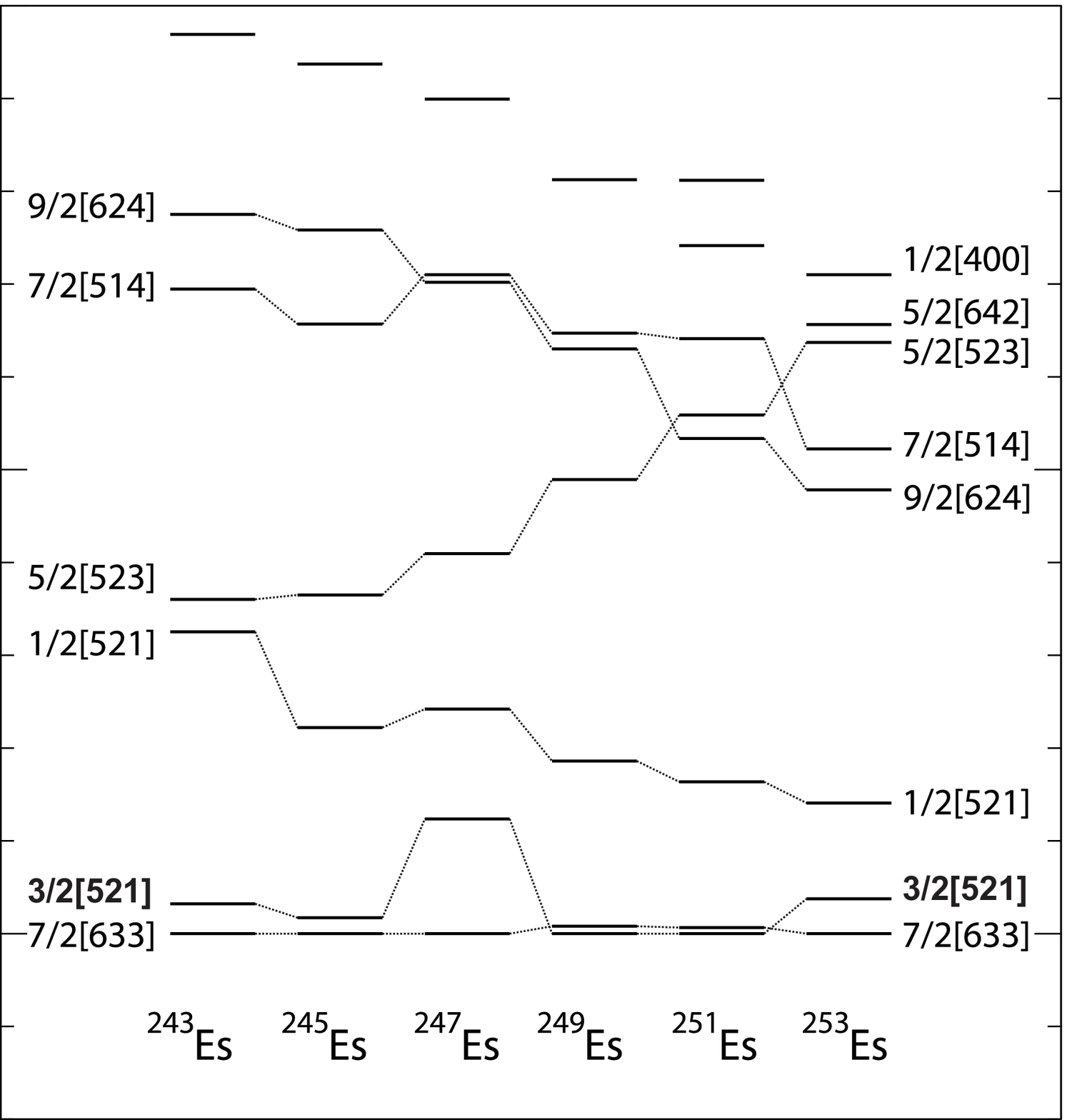,width=1.945in}
\end{center}
\caption{Quasiparticle energies calculated for the
UNEDF1$^{\textrm{SO}}_L$ parameter set\protect\cite{shi13},
and for the $N=151$ isotones (left) and Es ($Z=99$) isotopes (right).}
\label{fig2}
\end{figure}

Figure~\ref{fig2} (left) shows the calculated quasiparticle spectra
for $N=151$ isotones from $Z=94$ to 104. For these nuclei,
experimental data are available in this mass region~\cite{asai11}.
The observed ground states are dominated by the 9/2[734] orbital,
with the 5/2[622] state located at $\sim$200\,keV. In $^{245}$Pu and
$^{247}$Cm, our calculations predict ground states of 9/2[734]. For
heavier isotones, the 7/2[624] state becomes lower than the
9/2[734] state, but
only by several tens of keV. These two levels are right below the
$N=152$ shell gap. They are very close to each other, and change
order with increasing proton number. This indicates that (i)
there is a significant $N$-dependence of the single-particle and
quasiparticle energies and (ii) adjusting the spin-orbit force to obtain
better agreement for the $11/2^-$ state in $^{251}$Cf results in the $j_{15/2}$ and $g_{9/2}$
spherical shells being too close together.
The present calculations predict a 5/2[622] state which is too high
and underestimate the energies of the 7/2[624] and 1/2[620] states.

Comparing our results with those of Ref.~\refcite{parkhomenko05}, we
note the presence of low-lying 7/2[624] states there too. We also
note an overestimation of their 1/2[620] energies. The excitation
energy of the 1/2[620] orbital is related to the shell gap at
$N=152$. Its isospin dependence is very interesting -- we note that
our results reproduce the observed increase of excitation energy of
1/2[620] with increasing proton number~\cite{asai11}.

Experimental information on quasi-proton spectra are relatively
scarce\cite{herzberg08}. Close to $Z=100$, systematic spectra are
only observed for Es ($Z=99$) isotopes~\cite{hessberger05}.
Figure~\ref{fig2} (right) displays the calculated quasiparticle
spectra for $Z=99$ isotopes from $N=144$ to 154. One can see that for
the ground states the calculations predict two very close states,
3/2[521] and 7/2[633], which is consistent with experimental
observations~\cite{hessberger05}. Compared to the calculations
presented in Ref.~\refcite{hessberger05}, our results show a much
lower 3/2[521] orbital. This is because our calculations predict very
close 7/2[633] and 3/2[521] levels below the $Z=100$ shell opening,
whereas the WS calculations give them $\sim$200\,keV apart. The
7/2[514] state is also systematically observed in the Es isotopes.
However, our calculation fails to reproduce the increase of excitation
energy of this orbital with increasing neutron number. We also note
that in our calculations, the 5/2[523] states exhibit an increase
similar to that observed in experiment for the state attributed to
the 7/2[514] orbital. Further experimental and theoretical work is
needed for these odd-$Z$ nuclei.

\section{Conclusion}

We performed self-consistent Skyrme Hartree-Fock-Bogolyubov
calculations for nuclei in nobelium region. Deformations
$\beta_{2,4,6,8}$ were studied for even-even nuclei
$^{246,248,250}$Fm, $^{252,254}$No, and $^{256}$Rf. The quasiparticle
spectra for $N=151$ isotones and $Z=99$ Es isotopes were
systematically calculated by using the blocking approximation within
the Hartree-Fock-Bogolyubov method. Comparisons were made with the
available experimental data and WS calculations. It was found that
our results give similar deformations to those obtained within the WS
calculations. The observed quasi-particle energies are reasonably well
reproduced, especially for proton quasiparticle states.

\section*{Acknowledgments}
This work was supported in part by the Academy of Finland and
University of Jyv\"askyl\"a within the FIDIPRO programme, the
Centre of Excellence Programme 2012--2017 (Nuclear and Accelerator
Based Physics Programme at JYFL), the European Research Council through the SHESTRUCT project (grant agreement number 203481), and by the Office of Nuclear
Physics,  U.S. Department of Energy under Contract
No. DE-FG02-96ER40963.
We acknowledge the CSC - IT Center
for Science Ltd, Finland, for the allocation of computational
resources.

\bibliographystyle{ws-procs9x6}
\bibliography{ICFN5-proceedings-yshi}

\end{document}